# The concept of multifractal elasticity

Alexander S. Balankin [1,2]

*Instituto Tecnologico y de Estudios Superiores de Monterrey, Campus Estado de México, México 52926, Mexico*

**Abstract**

A new type of elasticity of random (multifractal) structures is suggested. A closed system of constitutive equations is obtained on the basis of two proposed phenomenological laws of reversible deformations of multifractal structures. The results may be used for predictions of the mechanical behavior of materials with multifractal microstructure, as well as for the estimation of the metric, information, and correlation dimensions using experimental data on the elastic behavior of materials with random microstructure.

## 1. Introduction

In the last decade the theory of materials with random (fractal or multifractal) microstructure has become an attractive topic in the mechanics and physics of solids [1-4]. The statistical properties of such microstructures are characterized by the spectrum of generalized dimensions $D_q$, where $-\infty < q < \infty$ (see, for example, Refs. [5,6]), also called Rényi dimensions and defined as

$$D_q = \lim_{r\to 0} \frac{I_q(r)}{\ln r},$$
$$I_q(r) = -\frac{1}{q-1} \ln\left(\sum_{i=1}^{N(r)} P_i^q(r)\right),$$
$$\sum_{i=1}^{N(r)} P_i(r) = 1. \qquad (1)$$

Here $I_q$ is the generalized entropy of order $q$, and $P_i(r)$ is the probability that a point of the structure considered lies in box (cell) number $i$ of the covering network with box size $r$. In general $D_q$ differs from the topological dimension $d_T$ of the multifractal structure embedded in $d$-dimensional Euclidean space. In particular, $D_0$ is equal to the metric dimension of the random structure evaluated by means of a box-counting algorithm, $D_F$, also called fractal dimension [6]. The generalized dimension of order $q = 1$ is equal to the information dimension $D_I$, which is associated with the information (Shannon) entropy $I_S = I_{q=1}$ [5]. The latter, as it was shown in Refs. [7,8] (see also Ref. [9]), decreases due to the elongation of the multifractal structure as

$$I_S(r) - I_{S0} \propto r^{D_I}. \qquad (2)$$

The generalized dimension $D_2$ is equal to the correlation integral exponent $D_C$, also called correlation dimension, which was introduced by Grassberg and Procaccia [10] as the exponent of a power-law correlation integral

[1] Formerly at the Dzerzhinski Military Academy, Moscow, Russian Federation.
[2] E-mail: balankin@servdgi.cem.itesm.mx.

$$C(r) = \lim_{N\to\infty} \frac{1}{N(N-1)} \sum_{i,j=1}^{N} \Theta(r - | r_i - r_j |) \propto r^{D_C}, \quad (3)$$

where $\Theta$ is the Heaviside function. The integral $C(r)$ counts the number of pairs of points (with coordinates $r_i$ and $r_j$) such that $r < | r_i - r_j |$ and thus governs the scaling of the internal energy of the deformed multifractal [7, 11]. Note that the generalized Rényi dimensions satisfy the general inequality

$$D_{q'} \leqslant D_q, \quad \text{for } q' > q, \quad (4)$$

the equality being obtained in the case of uniform sets (monofractals), i.e., such that the probability measure is constant, $P_i \equiv 1/N(r)$, and the generalized dimension $D_q$ equals the metric (fractal) dimension $D_F \leqslant d$ for all $q$ [5].

Aerogel, colloidal aggregates, polymers, rocks, some types of composite materials, porous media, etc. have a multifractal structure in a wide (but bounded) range of spatial scales $L_0 < L < L_M$ [4,5,11–13]. The elastic behavior of such materials generally differs from those for the classic elastic continuum even in the zeroth long-wave limit [3,4,12]. Therefore coupled stress theories of the elasticity of media with microstructure, such as the Cosserat continuum, micromorphic media, etc., which all in the zeroth long-wave approximation are equivalent to the classical model of an elastic continuum [14], cannot describe the elastic behavior of materials with multifractal microstructure.

In practice, for modeling elastic behavior of materials with multifractal microstructure the concepts of energetically and entropically derived elasticity are broadly used [15–24]. Moreover, in the first case the elastic force may be due to interatomic interactions (crystals, amorphous materials, etc.), or result from the spring-like behavior of elements of the (micro) structure (foams, some structural composites, etc.). The corresponding theories of elasticity are based on different phenomenological laws (experimental facts) and lead to different types of constitutive equations.

The classical theory of elastic continuum is based on two experimentally established facts [25]:

(1) Hooke's law, according to which the strain $\varepsilon_{ij}$ is proportional to the applied stress $\sigma_{ij}$, i.e., $\sigma_{ij} = E\varepsilon_{ij}$, $i, j = 1, 2, \ldots, d$, where $E$ is the elastic modulus; and

(2) Poisson's effect of lateral deformations (transverse strains) without corresponding stresses, $\varepsilon_{jj} = -\nu_{ij}\varepsilon_{ii}$ $(j \neq i)$, $\sigma_{jj} = 0$, where $\nu_{ij}$ is the tensor of Poisson's ratios.

In practice, instead of these two postulates the generalized form of Hooke's law is used. According to this law Poisson's effect is a consequence of the symmetry of the stress and strain tensors [26]. For an elastically isotropic $d$-dimensional solid the generalized Hooke's law has the form

$$\sigma_{ij} = \frac{\partial \Phi}{\partial \varepsilon_{ij}} = B \sum_{i}^{d} \varepsilon_{kk}\delta_{ij} + 2G \sum_{ij}^{d} (\varepsilon_{ij} - \tfrac{1}{3}\varepsilon_{kk}\delta_{ij}), \quad (5)$$

or

$$\varepsilon_{ii} = \frac{\partial \Phi}{\partial \sigma_{ii}} = E^{-1} \left( \sigma_{ii} - \nu \sum_{j=1}^{d-1} \sigma_{jj} \right).$$

It should be emphasized, however, that, whereas the formula $\sigma_{ij} = \partial\Phi/\partial\varepsilon_{ij}$ is a general relation of thermodynamics, the inverse formula $\varepsilon_{ij} = \partial\Phi/\partial\sigma_{ij}$ is applicable only if Hooke's law is valid.

The linear theory of random elastic media obeying generalized law (5) is well developed. A particularly successful method is that of the so-called self-consistent scheme (also autocoherent scheme). This method leads to effective elastic moduli, the aim of many investigations. It was shown by Kröner [27], that the self-consistent results are rigorous for the so-called perfectly disordered media obeying the generalized law (5).

The basic postulates of classical molecular theories of rubber-like elasticity are [28]:

(1) Intermolecular interactions are independent of the configuration and thus independent of the extent of deformation and therefore play no role in deformations at constant volume and composition.

(2) The Helmholtz free energy $\Phi$ of the molecular network is separable, i.e., $\Phi = \Phi(T, V, N) + \Phi_e(\lambda_i)$, where $\lambda_i = l^i/l_0^i$ ($l_0^i$ and $l^i$ are the sample dimensions in the $i$-directions before and after the application of the stress, respectively); $T$ is the absolute temperature, $V$ is the volume, and $N$ is the number of elementary particles (elements of the network). In particular, the classical theory of rubber elasticity, which was devel-oped in the 40s independently by a number of workers [29-32], is based on (1) the assumption that the chains of a polymer network obey Gaussian statistics, and (2) the assumption of incompressibility of elastomers. Under these assumptions the entropy of a deformed network changes as

$$\Delta S = -AT\left(\sum_{i=1}^{d} \lambda_i^2 - 3\right), \tag{6}$$

where $A$ is a constant which is determined by the topological properties of the structure [28]. In the case of uniaxial tension (compression) the dependence of the stress $F_1$ (per unit cross-section area of the sample in the initial state) on the relative strain in the direction of the applied force $\lambda_1$ is

$$F_1 = \tfrac{1}{3}E(\lambda_1 - \lambda_1^{-2}),$$
$$\lambda_2 = \lambda_3 = \lambda_1^{-0.5} \quad (\lambda_1\lambda_2\lambda_3 \equiv 1). \tag{7}$$

Table 1
Comparison of Poisson's ratio $\nu$, calculated using the analytical relation (13), with the computed values, based on two-dimensional elastic random networks, and with experimental data for aerogel $SiO_2$, strongly twisted nondeformable polymer filament, and rubber

| Properties | Two- dimensional random network of size $L \times L$ near the percolation threshold ($\xi_C$ = correlation length) | | Aerogel $SiO_2$ | Strongly twisted nondeformable filament | Rubber |
|---|---|---|---|---|---|
| | $L/\xi_C \to \infty$ | $L/\xi_C \to 0$ | | | |
| Connectedness of random network | Elasticity of network is determined by dangling bonds | Bonds, determining elasticity of network are multiduplicated | Fractal cluster | Monomer | Polymer network |
| Fractal dimension of elastic backbone $d_F$ | Dimension of geodesic line, $d_F = d_{min}$, $d_{min} = 1.1 \pm 0.02$ [41] | Dimension of red bonds, $d_F = d_{rb}$, $d_{rb} = 3/4$ [41] | $d_F$, measured by small-angle neutron scattering and molecular adsorption [42]: $2.3 \pm 0.1$ | Dimension of self-avoiding random walk: $d_{s-a} = 2$ [41] | $d_F$ $= 3.00 \pm 0.04$ [43, 44] |
| $\nu_F$, eq. (13) | $0.1 \pm 0.01$ | $-1/3$ | $0.15 \pm 0.05$ | 0 | $0.50 \pm 0.02$ |
| Poisson's ratio (results of numerical simulation and experimental data) | $0.08 \pm 0.04$ [16] | $-0.33 \pm 0.01$ [16] | $0.12 \pm 0.08$ [45] | 0 [46] | $0.50 \pm 0.01$ [28] |

Here $E$ is the Young modulus, which is proportional to the temperature [28].

The spring-like elasticity [33] is related to the empirical relation between the relative strain $\lambda_1$ and the applied force

$$F = E(\lambda - 1), \tag{8}$$

valid for springs, long polymer chains, elastic foams, etc. [28,33,34]. It should be noted that the origin and mechanism of lateral deformations of materials which show spring-like elasticity (for example, elastic foams [33,34]), differ from those for the elastic continuum (Poisson's effect), as well for elastomers (condition of incompressibility), and have a purely geometrical nature.

In this article we suggest a more general type of reversible deformations of elastic random structures, that is the multifractal elasticity, and show that rubber-like elasticity, superelasticity, and spring-like elasticity may be interpreted as special cases of multifractal elasticity.

## 2. Basic postulates

In developing the theory of multifractal elasticity we also start with two postulates (laws), which we establish on the basis of the analysis of experimental data and results of computer simulations of the elastic behavior of multifractal structures:

(1) When an external force $F$ is applied to an elastically isotropic multifractal object, deformations occur mostly on a length scale beyond a certain characteristic length, which changes with increasing stress. Thus, the presence of an external stress leads to the appearance of the unique new characteristic scaling length $L_F$.

The initial morphology of a multifractal can either be characterized by one or more length scale parameters $L_i$, or not have any at all. In the first case the $L_F$ may have physical significance such as the characteristic dimension of blobs in polymer, the characteristic size of cells or the mean distance between inclusions in a composite material, the characteristic radius of correlations in a random network and aerogels, etc. As it follows from the law postulated above, only one of the set of scale parameters $L_i$ (or their invariant combination) depends on the external forces. If the initial multifractal structure has no scale parameters, then the physical meaning of $L_F$ is the characteristic length above which deformations occur [3].

A regular lattice may be interpreted as a limiting case of a multifractal structure for which all $D_q$ equal the topological dimension of the lattice $d_T$. In this case $L_F$ is simply the length of interatomic bonds. Notice, however, that the mechanism of elasticity discussed here is not valid for the elastic continuum and common crystal materials obeying in the zeroth long-wave limit the generalized Hooke's law (5).

If the proposed law is valid, from the second law of thermodynamics it follows that the force obeys

$$F = \left(\frac{\partial \Phi}{\partial L_F}\right)_T = \left(\frac{\partial U}{\partial L_F}\right)_T - T\left(\frac{\partial S}{\partial L_F}\right)_T. \tag{9}$$

The first term on the right hand side of Eq. (9) is evidently the energy component of the internal forces, and the second term is the entropic component.

It is well known that in the zeroth long-wave approximation, at not very high frequencies, the differences between models of media with microstructure disappear [14]. Therefore, for a static theory of elasticity of multifractals, which is considered below, it is possible (and sufficient) to formulate the first law in terms of the relative deformations with the scaling parameter

$$\lambda_F = \frac{L_{F_1}}{L_{F_2}}, \tag{10}$$

[3] It should be remarked that the idea that the presence of an external stress introduces a new length scale in the problem of elastic deformations of random structures was invoked by Pincus [35] and de Gennes [36] to study the elasticity of macromolecules, and by Webman [17] to construct a heuristic picture of elastic deformations of monofractal structures.

where $L_{F_1}$ and $L_{F_2}$ are the longitudes of the scaling length $L_F$ corresponding to $F = F_1$ and $F = F_2$, respectively. We emphasize, however, that instead of the nonlocal theories of elasticity considered in Ref. [14], which all in the zeroth long-wave approximation are equivalent to the classical elastic continuum model, the zeroth long-wave approximation in our theory differs from the elastic continuum model, because of the different basic laws.

(2) During a reversible deformation a multifractal structure retains scaling in the mass density, i.e., the relation which governs the change in the mass density $\rho$ during the elastic deformation of the multifractal is similar to the relation which governs the change in the mass density because of the geometric changes in the dimensionalities of structure, namely,

$$\frac{\rho(F_1)}{\rho(F_2)} = \left(\frac{L_{F_1}}{L_{F_2}}\right)^{-\alpha}_{T=\mathrm{const}} = \lambda_F^{-\alpha}, \quad \alpha = d - D_F. \tag{11}$$

Notice that this relation is the generalization of the condition of incompressibility (second relation in Eq. (7)). It is obvious, that relation (11) is not valid for the elastic continuum ($D_F = d$) with Poisson's effect.

Instead of the condition of the homeomorphism of reversible deformations, which is used in the mechanics of the continuum [37], we assume that there are no changes in the metric, information, and correlation dimensions of the elastically isotropic multifractal during its reversible deformation" [4].

It follows from Eq. (11) that under uniaxial tension (compression) the change in the dimensionality of the multifractal structure in the direction of the external force $F_x$, which is $\lambda_x = L_x/l_x$, is accompanied by a change in its lateral dimensionalities in the orthogonal directions of the surrounding $d$-space with $\lambda_i = L_i/l_i$, where $i = 1, 2, ..., d-1$. The lateral deformations $\lambda_i$ are related to $\lambda_x = \lambda_F$ as

$$\lambda_i = \lambda_\perp = \lambda_x^{-\nu_F} = \lambda_F^{-\nu_F}, \quad i = 2, 3, ..., d, \tag{12}$$

and thus we have $\alpha = 1 - (d-1)\nu_F$, so that

$$\nu_F = -\frac{\ln \lambda_\perp}{\ln \lambda_F} = \frac{d_F}{d-1} - 1. \tag{13}$$

Therefore, if the law (11) is valid, the transverse deformation exponent of an elastically isotropic multifractal is defined uniquely by its metric dimension.

At a first glance, it is surprising that lateral deformations are independent of the geometry of the deformed multifractal structure, whose transverse deformation exponent $\nu_F$ is determined only by its metric dimension. However, it should be recognized that similar phenomena (power law distributions of strains and stresses with exponents which are functions of the Poisson ratio) are common within singular problems of the classic linear theory of elasticity (see, for example, Refs. [39,40]). It should be noted that precisely the change in the distribution of internal stresses due to the spatial inhomogeneity of the multifractal structure gives rise to its lateral deformations under external loading. Notice that the exponent of lateral deformations $\nu_F = -\ln \lambda_\perp / \ln \lambda_F$ is equal to the Poisson ratio

$$\nu = -\sqrt{(\lambda_i^2 - 1)/(\lambda_j^2 - 1)}$$

only in the limit of infinitely small strains $\varepsilon = \sqrt{|\lambda_i^2 - 1|} \ll 1$.

The data of Table 1 demonstrate that the results of calculations by the analytical formula (13) agree well with the results of numerical simulations of the elasticity of two-dimensional percolation networks near the percolation threshold, as well as with the experimental data for aerogel $SiO_2$ (which were obtained in studies of the propagation of longitudinal and transverse elastic waves), rubber, and strongly twisted nondeformable polymer filaments.

It must be emphasized that for regular lattices $D_F = d_T$ and according to (11) we always have either the condition of incompressibility

$$\nu_F = \frac{1}{d-1}, \tag{14}$$

if the topological dimension of lattice $d_T$ equals the topological dimension of the enveloping Euclidean space $d$, or the absence of lateral deformations, i.e., $\nu_F = 0$, if $d_T = d - 1$. Hence, for materials obeying the generalized Hooke's law our conjecture (11) is not valid.

[4] It is clear that the homeomorphic deformations do not change the metric dimension of the structure. It was shown [38] that there are no changes in $D_F$, $D_I$, and $D_C$ of the multifractal after an affine transformation (deformation). Moreover, in the case of affine deformations relation (11) may be derived rigorously from the scaling properties of the multifractal structure for which our first postulate is valid.

Notice that Eqs. (12) and (13) are satisfied in the general case of an $n$-dimensional deformation of a multifractal structure obeying laws (9) and (11). For example, in the case of a biaxial extension (compression) of an elastic multifractal ($1 \leqslant d_F \leqslant 3$) in three-dimensional space ($d = 3$) it follows from Eq. (11) that $\lambda_z = \lambda_F^{-\nu_F}$, $\lambda_F = (\lambda_x \lambda_y)^{1/(1-\nu_F)}$, where $\nu_F$ and $D_F$ are still related by Eq. (13), which is satisfied also in the case of a triaxial deformation of an elastic fractal in three-dimensional space when $\lambda_F = (\lambda_1 \lambda_2 \lambda_3)^{1/\alpha}, \alpha = d - D_F = 1 - 2\nu_F$, and the law (11) is valid, i.e., $p \propto \lambda_F^{-\alpha}$. The pure shear can be represented as a biaxial loading under forces $F_1$ and $F_2$ such that there is no change in length along the direction of the second force.

## 3. Constitutive equations for elastic multifractals

Using the definitions of the generalized dimensions (1) and the scaling properties of the information entropy (2) and the correlation integral (3) with laws (9), (11), it is easy to show that the change in the thermodynamic entropy $\Delta S(\lambda_i) \propto \Delta I_S(\lambda_i)$ during the reversible deformation of an elastic multifractal in $d$-dimensional space can be represented in the form

$$\Delta S = -C_1 \left( \sum_{i=1}^{d} \lambda_i^{D_I} - d \right), \tag{15}$$

while the change in the internal energy $U(\lambda_i)$ is equal to

$$\Delta U = -C_2(\lambda_F^{\alpha_C} - 1), \quad \alpha_C = d - D_C, \tag{16}$$

where the parameters $C_1$ and $C_2$ do not depend on $\lambda_i$ and may be also determined for any detailed model of the structure.

Notice that for multifractals obeying Gaussian statistics and thus characterized by $D_I = 2$ [11,47] relation (15) reduces to the classic formula (6).

Substituting (15), (16) into (9), and using Eqs. (10), (12), and (13) we can obtain the relationships between the external force $F_i$ and the relative deformations $\lambda_i$ of an elastic multifractal structure. For example, in the case of uniaxial tension (compression) we have

$$\begin{aligned} F_1 = C_1\{ & D_I \lambda_1^{D_I - 1} \\ & - D_I[d_F - (d-1)] \lambda_1^{-D_I[D_F/(d-1)-1]-1} \\ & - (C_2/C_1)(d - D_C)\lambda_1^{d - D_C - 1}\}. \end{aligned} \tag{17}$$

According to the obvious condition $F_1(\lambda_i = 1) = 0$, it follows from Eq. (17) that

$$\frac{C_2}{C_1} = D_I \frac{d - D_F}{d - D_C} \leqslant D_F. \tag{18}$$

In the case of a monofractal structure all generalized dimensions equal the metric (fractal) dimension, i.e., $D_q = D_F = D_I = D_C$, and from Eq. (18) it follows that $C_2/C_1 = D_F$. In general, putting speculatively $C_2/C_1 = D_C$ in Eq. (18) we obtain the following relationship between the three dimensions of an elastic multifractal,

$$D_I(d - D_F) = D_C(d - D_C). \tag{19}$$

Notice that this conjecture, as well as equality (18), is valid only for elastic multifractals obeying laws (9) and (11).

Using Eq. (18) we can rewrite Eq. (17) in the form

$$\begin{aligned} F_1 = C_1 D_I \{ & \lambda_1^{D_I - 1} - [D_F - (d-1)]\, \lambda_1^{-\nu_F D_I - 1} \\ & - (d - D_F)\lambda_1^{\alpha_C - 1}\}. \end{aligned} \tag{20}$$

The stress $\sigma_{11}$ is related to $F_1(\lambda_1)$ by the obvious equation $\sigma_{11} = F_1 \lambda_1^{1-\alpha}$, which allows for the change in the area of the $(d-1)$-dimensional cross section of a deformed fractal in the plane orthogonal to the direction of the external force $F$.

In the case of a monofractal structure using relation (13) we obtain the equation for $\sigma_{11}$ in the form

$$\begin{aligned} \sigma_{11} = & \frac{E}{1 + 6\nu_F + 4\nu_F^3} \\ & \times [(\lambda_1^{1+4\nu_F} - 1) - 2\nu_F(\lambda_1^{-1-2\nu_F^2} - 1)], \end{aligned} \tag{21}$$

where $E = (\partial\sigma_{11}/\partial\lambda_1)_T$ is the Young modulus. It is easy to see that within the limit of infinitesimally small strains, $|\varepsilon_{11}| = \sqrt{|\lambda_1^2 - 1|} \ll 1$, Eq. (21) leads to its classic counterpart, Eq. (5). Thus, for an elastic monofractal we have

$$C_1 = 2(1 + \nu_F)C_2 = (1 + 6\nu_F + 4\nu_F^3)E.$$

Similarly, we can derive the relations $\sigma_{ij}(\lambda_k)$ for an $n$-axial deformation of an elastically isotropic monofrac-tal in the $d$ space. The state of deformation for a biaxial extension, where the dimensions of the network are changed independently along two axes, is given by $F_1 = f(\lambda_1, \lambda_2), F_2 = f(\lambda_2, \lambda_1), \lambda_3 = \lambda_F^3/\lambda_1\lambda_2$. The pure shear state is essentially a biaxial loading under the stresses $\sigma_{11}$ and $\sigma_{22}$ such that there is no change in the length along the second direction and in the volume, i.e., $\lambda_2 = 1, \lambda_3 = \lambda_1^{-1}$ $(\lambda_1\lambda_3 = 1)$. In the limit of infinitesimally small strains $\varepsilon_{ij} \ll 1$ the set of these relations may be represented by the classical form of the generalized Hooke's law (5). This allows us to find the relationships between the elastic moduli, i.e., Young's modulus $E = (\partial\sigma_{11}/\partial\varepsilon_{11})_T$, the shear modulus $G$, and the bulk modulus $B$, and the Lamé coefficients $\lambda, \mu$ of a monofractal structure,

$$G = \frac{E(d-1)}{2d_F}, \quad B = \frac{E}{d(d-d_F)},$$
$$B = \lambda + \frac{2}{d}\mu, \tag{22}$$

which are derived in analogy to the derivation of the corresponding relationships in the classical theory of elasticity (see, for example, Refs. [25,26]). Notice that relations (22) differ from those which were conjectured for elastic monofractals by Bergman and Kantor [16]. At the same time, substituting Eq. (13) in Eqs. (22) we obtain expressions that for $d = 2$ and $d = 3$ are identical to those for two- and three-dimensional elastically isotropic continua [5].

Notice, that for long polymer chains characterized by $D_F = d = 2$ and $D_I = 1$ [48], the basic relation of spring-like elasticity (8) results from Eqs. (17), (18).

## 4. Multifractal models of rubber-like and super-elasticity of polymers

It is easy to see that the classical formula of the rubber elasticity (7) may be derived from (17), (18) for multifractals obeying $D_F = d = 3$ and $D_I = 2$. In fact, however, calculations based on Eq. (7) with the value of $E$ adjusted by fitting are in reasonable agreement with the experiments only in the range of relatively small strains ($\lambda_i < 1.2$) [28,33]. When $1.2 \leqslant \lambda_1 \leqslant 2$, the graph of relation (7) usually lies above and for $\lambda_1 > 2$ well below the experimental curve $F_1(\lambda_1)$, which in the range $\lambda_1 > 4$ obeys asymptotically

$$F_1 \propto \lambda_1^2. \tag{23}$$

Traditionally, relation (7) is refined by phenomenological modifications of the entropic theory, or by using empirical models of the elastic potential (see Refs. [28,33,49,50], and references therein). Apart from the large number of and not always clear physical meaning of the matching parameters, the main shortcoming of both such modifications of the classic entropic theory of rubber elasticity and the empirical models of the (hyper)elastic potential is that different values of the same elastic parameters such as $E$ are needed to describe the experimental data obtained under different loading conditions, and also to describe the same data, but within the frameworks of different modifications of the entropic theory or the elastic potential. This is due to the indeterminacy of the absolute values of the elastic moduli and the relationships between them, which (within the framework of the classical theory of rubber elasticity) do not satisfy the expressions for the elastic continuum even in the limit of infinitesimally small strains.

Elastomers are known to have a fractal or multifractal microstructure [43,44,46,48]. Therefore, it is natural to describe the rubber elasticity of polymers by using the concept of multifractal elasticity discussed above.

In the general case the generalized dimensions of polymer networks swelled in a good solvent are within the range $2 < D_q \leqslant 3$ [43,44]. Assuming in the first approximation that $D_F = D_I = D_C$ and substituting Eq. (13) into Eq. (17), we obtain the relationship between the nominal stress $F_1$ and the strain factor $\lambda_1$ in the case of uniaxial tension (compression) of an elastomer,

$$F_1 = \frac{E}{1+6\nu_F+4\nu_F^3}[\lambda_1^{1+2\nu_F} - 2\nu_F\lambda_1^{-1-2\nu_F(1+\nu_F)} - (1-2\nu_F)\lambda_1^{-2\nu_F}]. \tag{24}$$

Notice that this relationship differs from (7) even in the limit of an incompressibly deformed material ($\nu_F \equiv 0.5$), when Eq. (24) reduces to the formula

[5] This result is very surprising, because the nature and mechanisms of lateral deformations for an elastic continuum (Poisson's effect) and for the fractal structure (law (11)) are different.

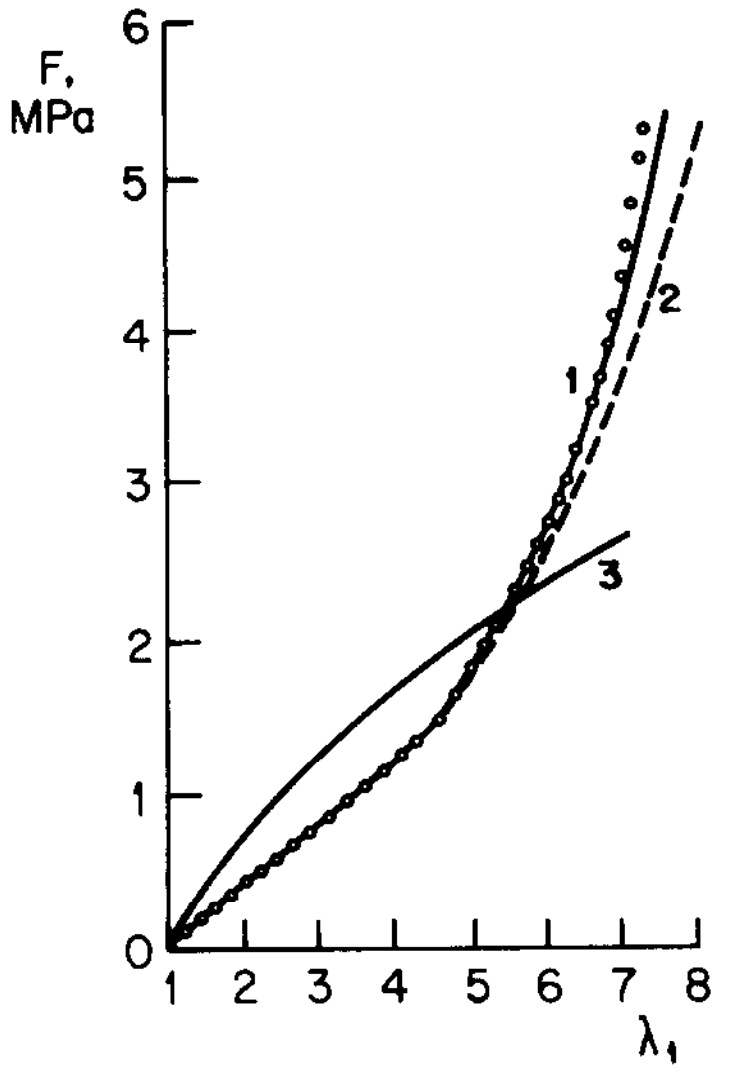

Fig. 1. The stress-strain behavior for rubber under uniaxial extension: (1) calculations by Eq. (25) with $E = 0.36$ MPa ($\nu_F \equiv 0.5$); (2) calculations by Eq. (24) with $E = 0.34$ MPa and $\nu_F = 0.48$; (3) calculations by classic formula (7) with $E = 0.4$. The points represent experimental data from Ref. [28].

$$F_1 = \frac{E}{4.5}(\lambda_1^2 - \lambda_1^{-2.5}), \qquad (25)$$

obeying the asymptotic (23).

It is clear from the graphs in Fig. 1 that the calculations based on Eqs. (24), (25) are in excellent agreement with the experimental data for rubber without any adjustment of parameters (except $E$) right up to $\lambda_1 = 7$.

The nonlinear stress-elongation asymptotic for superelastic networks [46],

$$\sigma_{11} \propto \lambda_1^{1/3}, \qquad (26)$$

may be also derived within the framework of Eqs. (17), (18) in the case of multifractal structures whose generalized dimensions obey the relation

$$D_F + D_l - d = \tfrac{1}{3}. \qquad (27)$$

In particular, for multifractals, which are characterized by $D_F = 3$ and $D_l = 1/3$, the superelastic behavior (26) has a purely entropic nature.

Thus, proper regard for the real topology allows an adequate description of the behavior of reversibly deformed materials with multifractal microstructure. Moreover, using the results of the experimental investigation of the elastic behavior of materials with multifractal microstructure (which obey laws (9), (11)) it is possible to calculate their metric, information, and correlation dimensions by means of Eqs. (13), (17), and (18).

## Acknowledgement

The author is grateful to G. Cherepanov for useful discussions. The courtesy and support of the staff and faculty of the Materials Research Center at the Monterrey University of Technology, and especially A. Bravo, are gratefully acknowledged.